\documentclass[twoside]{article}
\usepackage{fleqn,espcrc2}
\usepackage{graphicx}
\usepackage{here}

\newcommand{\AmS}{{\protect\the\textfont2
    A\kern-.1667em\lower.5ex\hbox{M}\kern-.125emS}}
\evensidemargin- 0.cm
\def\beq{\begin{equation}}
\def\eeq{\end{equation}}
\def\bea{\begin{eqnarray}}
\def\eea{\end{eqnarray}}
\def\bq{\begin{quote}}
\def\eq{\end{quote}}

\def\nnb{\nonumber}
\def\ga{\left(}
\def\dr{\right)}

\def\rar{\rightarrow}

\def\nnb{\nonumber}
\def\la{\langle}
\def\ra{\rangle}

\def\ba{\begin{array}}
\def\ea{\end{array}}

\def\als{\alpha_s}

\def\gg2{ \la\alpha_s G^2 \ra}
\def\gg3{g^3f_{abc}\la G^aG^bG^c \ra}
\def\ggg4{\la\als^2G^4\ra}

\title{\bf{\boldmath
{\Large Can the $\gamma\gamma$ processes reveal the nature of  the $\sigma$ 
meson ? } }}
\author{
G. Mennessier \address {\footnotesize Laboratoire
de Physique Th\'eorique et Astroparticules, Universit\'e
de Montpellier II, Case 070, Place Eug\`ene
Bataillon, 34095 - Montpellier Cedex 05,
France,
{Email: gerard.mennessier@lpta.univ-montp2.fr} },
P. Minkowski \address {\footnotesize Institut f\"{u}r Theoretische Physik, University of Bern,  Sidlerstrasse 5, 
CH-3012 Bern, Switzerland, 
Email: mink@itp.unibe.ch},
S. Narison
\address {\footnotesize Laboratoire
de Physique Th\'eorique et Astroparticules, Universit\'e
de Montpellier II, Case 070, Place Eug\`ene
Bataillon, 34095 - Montpellier Cedex 05,
France,
Email: snarison@yahoo.fr} 
 and 
W. Ochs \address {\footnotesize  Max-Planck Institut f\" ur Physik, 
D 80805 Munich, F\"ohringer Ring 6, Germany,
  {and Laboratoire
de Physique Th\'eorique et Astroparticules, Universit\'e
de Montpellier II, Case 070, Place Eug\`ene
Bataillon, 34095 - Montpellier Cedex 05,
France, Email: wwo@mppmu.mpg.de }} \\
}

\begin{document}

\pagestyle{empty}
\pagestyle{plain}
\begin{abstract}
\noindent
We reanalyse the $\gamma\gamma$ scattering data and conclude that in
 the
mass region below 1 GeV the cross section for 
$\gamma\gamma\to \pi^0\pi^0$ can be largely explained by
the one pion exchange process with $ \pi\pi$ rescattering. 
The radiative width of the $\sigma$ is estimated and a model dependent
separation into contributions from direct $\gamma\gamma$ decay and decay
through rescattering is obtained.
We confront these findings 
with QCD spectral
 sum rule (QSSR) predictions and conclude that the $\sigma$ can have a
 large gluonium component in its wave function.
\end{abstract}
\maketitle
\section{Introduction}
 \par
Understanding the nature of scalar mesons in terms of quark and gluon
constituents
is a long standing puzzle in QCD \cite{MONTANET}. One might expect that
the decay rate of these mesons into two photons could provide an
 important 
information
about their intrinsic composite structure. The problem here is that
 some 
states , discussed intensively at present, are very broad
($\sigma$ and $\kappa$ mesons), others are close to an inelastic
 threshold
($f_0(980)$, $a_0(980)$), which makes their interpretation more
 difficult. 
Besides the interpretation within a $q\bar q$ model
\cite{MONTANET,MORGAN,BN,SNG,KLEMPT,OCHS,SN06}
or unitarized quark model \cite{TORN,BEVEREN},
 also the possibility of
tetraquark states \cite{JAFFE1,BLACK,SNA0,ACHASOV} (and some other related
 scenarios: meson-meson molecules \cite{ISGUR,BARNES}, 
meson exchange \cite{HOLINDE}) is considered. 
In addition, a gluonic meson is expected in
the scalar sector, according to QCD spectral sum rules (QSSR)
\cite{NSVZ,SNG0,SNG1,VENEZIA,SNG,SN06} 
and lattice QCD \cite{PEARDON,MICHAEL}. Such a
state could mix with the other $\bar qq$  mesons 
 \cite{BN,SNG,OCHS,CLOSE,aniso}. 
Among the light particles, the 
$\sigma$ meson could be such a gluonic 
 resonance, that can manifest itself in some effective linear sigma models
 \cite{LANIK,ZHENG} or contribute to the low-energy constants at ${\cal O}(p^4)$ of
the QCD effective chiral Lagrangian \cite{PICH}.

The existence of glueballs/gluonia is a characteristic prediction of
 QCD and some
scenarios have been developed already back in 1975 \cite{MIN}. 
Today, there is agreement that such states exist in QCD and the
 lightest state has
quantum numbers $J^{PC}=0^{++}$. Lattice QCD calculations in the
 simplified 
world without quark pair creation (quenched approximation) find the
lightest state at a mass around 1600 MeV \cite{PEARDON}. These findings
 lead to the construction of models
where the lightest glueball/gluonium mixes with other mesons in the
 nearby mass range
at around (1300-1800) MeV (see, for example, \cite{CLOSE}). 
However, recent results beyond this approximation    
\cite{MICHAEL} suggest that the lightest state with a large gluonic
 component
is rather in the region around 1 GeV and, therefore, a scheme based on 
mixing involving only meson states
with mass higher than 1300 MeV could be insufficient to
represent the gluonic degrees of freedom in the meson spectrum.

The approach based on QSSR \cite{SVZ,SNB} has given quantitative
 estimates of the
mass of  glueballs/gluonia and also of some essential
 features of its
branching ratios. The mass of the lightest scalar gluonium is \cite{SNG0,VENEZIA,SNG,SN06} :
\beq
M_{0^{++}} \approx  (750\sim 1000)~{\rm  MeV}~,
\label{eq:scalarmass}
\eeq
with a corresponding total width ranging from 
300 to 1000 MeV (see section \ref{qssr}).
Some large mixing of this gluonium is expected with the nearby isoscalar $q\bar q$ states
  resulting in the physical states  $\sigma$ and
 $f_0(980)$\cite{BN,SNG}.


In phenomenological studies of $\pi\pi$ interactions, a broad object has been
identified. Because of the large width and the presence of other resonances
overlapping the identification is not straightforward. This is also
reflected in the record of the Particle Data Group refering previously to
$f_0(400-1200)$ and now to $f_0(600)$ or $\sigma$. Definitive studies have
been carried out in elastic $\pi\pi$ scattering where energy independent
phase shift analyses allow the reconstruction of the unitary amplitude.  The
S-wave isoscalar scattering phase shift $\delta_0^0$ rises slowly and moves
through 90$^\circ$ around 850 MeV \cite{CERN-MUNICH} and
 it continues its rise also above the
inelastic $K\bar K$ threshold in the observed range up to 1800 MeV.
Ambiguities occuring in the phase shift analysis \cite{EM,CM2}
have been successively resolved \cite{KPY,ochs06} by including data from 
$\pi\pi$ charge exchange \cite{GAMS}
 and the resulting
complex amplitude nearly completes a full circle in the Argand diagram.  In
between, two narrow states, $f_0(980)$ and $f_0(1500)$ are clearly
established and are superimposed to the slow movement of the ``background
amplitude''. 

In the K-matrix fits to the $\pi\pi$ elastic scattering data up
to the highest available energy [(500-1800) MeV], a pole is found in the
S-wave amplitude with a large imaginary part which corresponds to a state of
large width \cite{CERN-MUNICH,ESTABROOKS,AMP}: 

\beq 
\Gamma\approx M\approx 1~{\rm GeV}~, \label{sigmamass} 
\eeq 
though this value may depend on the
treatment of the other resonances \cite{aniso,OCHS,ochs06,mennessier}.

The broad object,  so defined as $\sigma$, has been identified and
singled out as a 
``left over state'' in a phenomenological classification of the low lying 
scalar meson spectrum
 into $q\bar q$ multiplets and has 
therefore been related to the lightest glueball \cite{OCHS}.
The appearence of this state in gluon rich processes was considered 
in
support of this hypothesis. In this analysis, results from elastic and
inelastic $\pi\pi$ scattering as well as from $D$, $B$ 
and $J/\psi$ decays have been considered \cite{OCHS,mo2,minkB}.

Because of its large width, which is of the order of its mass, 
the $\sigma$, as a particle, is standing out
and predictions involving its
mass and width are of a particular concern.
The mass parameter in (\ref{sigmamass}) is close to the value where the 
observed S-wave phase shift goes through  $90^\circ$ as in a simple 
Breit Wigner form without background. 
This parameter also depends on the mass range taken into account 
and analyses which do not include
the high mass tale of the S wave spectrum above 1 GeV
tend to yield smaller values for this mass.  

In general, the complex resonance 
self-energy $\Sigma=M-(i/2) \Gamma$ 
is energy dependent
and determines the zero's and poles of the S-matrix
amplitude joining appropriate sheets in the cut s-plane. 
In recent determinations, where analyticity and unitarity properties 
are used to continue the amplitude into
the deep imaginary region,  values: 
\bea
M_\sigma &=& 441-{\rm i} 272 ~{\rm MeV}~  \cite{leutwyler} \nnb\\
&=& 489-{\rm i}264~{\rm MeV}~  \cite{YND} 
\label{sigma}
\eea
have been obtained. We assume here that these poles refer
 to the 
$\sigma$ as defined above.
We use below 
the values in eq. (\ref{sigmamass}) under the assumption that, in all
directly observable experimental numbers, this is a good approximation,
leaving aside the actual extrapolation to complex kinematic variables,
 which needs theoretical assumptions or approximations.

In the above mentioned prediction of the glueball/gluonium mass from
 QSSR, a
narrow width approximation or a Breit Wigner form parametrization 
of the spectral function  has been employed. 
So we compare it
with the mass in (\ref{sigmamass}),  obtained in a similar approximation from the resonance
amplitude, which is closer to the mean of the observed
mass spectrum of the amplitude squared $|T_0^0|^2$. In this  sense
the prediction for the
mass of $\sim 1 $ GeV \cite{SNG,VENEZIA} agrees with the observed
 $\sigma$ ``Breit Wigner mass".

We are now coming back to the problems with the $\gamma\gamma$ width.
A recent analysis of the $\gamma\gamma\to \pi\pi$ processes
\cite{PENNINGTON} has extracted:
\beq
\Gamma (\sigma\rar\gamma\gamma)\simeq (4.1\pm 0.3)~{\rm keV}~,
\label{widthpenn}
\eeq
for the $\sigma$ mass obtained in \cite{leutwyler}.
The $\gamma\gamma$ width looks fairly high compared with most of  the
 available theoretical 
estimates based on QCD dynamics. \\
This result was interpreted in \cite{PENNINGTON} as
disfavouring a gluonic  and tetraquark  nature
which is expected to have a small coupling  to $\gamma\gamma$
 \cite{VENEZIA,BN,SNG,ACHASOV,BARNES,SNA0}.
In the following, we shall reconsider the analysis  of the
 $\gamma\gamma\to \pi\pi$
process in the low energy region below 1 GeV, where we conclude that it
 is 
dominated by the coupling of the photons to charged pions and their
 rescattering, which 
therefore can hide any 
direct coupling of the photons to
the scalar mesons. 
\section{Analysis of $\gamma\gamma$ scattering data}
A striking feature of the low energy $\gamma\gamma$ scattering is the
difference of cross sections for the charged and neutral $\pi\pi$ final
states: the charged final state is produced with a rate about an order
 of
magnitude larger than the neutral final state which can be due to
 the
contribution from the one-pion-exchange Born term in the process
$\gamma\gamma\to\pi^+\pi^-$. In the process $\gamma\gamma\to
 \pi^0\pi^0$,
the photons cannot couple ``directly'' to $\pi^0\pi^0$ but through
 intermediate charged
pions and subsequent rescattering with charge exchange. 
In fact, in specific field theory models, this is the
dominant mechanism for the process with neutral pions and, in the following,
 we shall discuss  two such examples. 
\subsection{Chiral perturbation theory}
In the effective chiral Lagrangian approach based on $SU(2)_L\times
 SU(2)_R$
symmetry, the interactions between pions and photons are given in terms
 of
parameters $m_\pi,f_\pi$ and $e$. 
To one-loop accuracy, the cross-section can be written in the
 factorised form 
\cite{donoghue}: 
\bea
\sigma(\gamma\gamma\to \pi^0\pi^0)=&& \ga\frac{2\alpha^2
 q}{8\pi^2\sqrt{s}}\dr\ [1+\frac{m_\pi^2}{s}f(s)]\nnb\\
&\times&\sigma(\pi^+\pi^-\to \pi^0\pi^0),
\label{chpth}
\eea
with $4q^2=s-4 m_\pi^2$ and:
$
f(s)= 2 [\ln^2(z_+/z_-)-\pi^2] + {m^2_\pi\over
 s}[\ln^2(z_+/z_-)+\pi^2]^2$,
where $z_\pm \equiv (1/2)\ga1\pm (1-4m^2_\pi/s)^{1/2}\dr$.  For large
 $s$ or/and in the chiral limit $m^2_\pi\rar 0$,
the chiral correction behaves as $(m_\pi^2/s)\ln^2(s/m^2_\pi)$, which
 is finite and tends to zero.
The $\pi^+\pi^-\to\pi^0\pi^0$ cross-section reads:
\beq
\sigma(\pi^+\pi^-\to\pi^0\pi^0)=(8\pi/9q^2)|T|^2~,
\eeq
and in terms of Isospin amplitudes and phase shifts 
$T_I=\sin \delta_Ie^{i\delta_I}$
\beq
T=T_0-T_2;\quad 
|T|^2=\sin^2(\delta_0-\delta_2)~.
\label{amp00}
\eeq
The 1-loop prediction meets the experimental data at around 500 MeV but
 is a bit
below at lower energy and above at the higher energies. In this
approximation, the amplitude is real and is strictly valid only in 
the threshold region. If one extrapolates Eq. (\ref{chpth}) based on
 one-pion
exchange towards higher energies, one 
obtains at 600 MeV with $|T|^2\approx 0.75$ 
a cross section of $\sim 9$ nb
which is near the experimental result of $\sim 12$ nb \cite{cball}. At
 this energy, the
correction term $(m_\pi^2/s)f(s)$ amounts to $\sim$10\% 
and the two cross sections in (\ref{chpth}) are almost proportional. 
An improved agreement with the data up to about 700 MeV 
has been obtained including  2-loop corrections \cite{bellucci}.
The changes to the 1-loop results amount to about 30\% and the
 important
role of one pion exchange is confirmed. 
\subsection{Analytic K-matrix model}
In  \cite{mennessier}, Mennessier introduced a model which
describes the $S$-wave $I=0$ $\pi\pi$ interaction using an analytic
  K-matrix approach with
two poles $\sigma$ and $f_0(980)$ resonances supplemented by
contributions induced by the 4-point $\pi\pi$ and
$K\bar K$ interaction vertices. The $\sigma$ pole is found at
$M_\sigma=500-i300$ MeV, not far away from (\ref{sigma})\footnote{A
 similar model has been also applied to  the $f_2(1270)$
resonance for the D wave in the mass range up to 1400 MeV.}.\\
In this approach, a subclass of bubble pion loop diagrams including
 resonance poles in the s-channel are resummed (unitarized Born). Coupling
 to photons is introduced through their coupling to charged pions, kaons
 and vector-mesons. \\
Like in the case of ChPT, the $\gamma\gamma\to \pi^+\pi^-$ process is
 dominated by the Born term, whilst, in
$\gamma\gamma\to\pi^0\pi^0$, the diagrams with the pion loop through
 one-pion exhange are most important below 1 GeV.
To 1-loop order, the expression of the $\gamma\gamma\to\pi^0\pi^0$ is
 similar with the one given by ChPT, where, to this order, and, at higher
 energies, the proportionality between the $\gamma\gamma\to\pi^0\pi^0$
 and the $\pi^+\pi^-\to \pi^0\pi^0$ cross-section continues to hold.  The
 free coupling parameters are fixed by a fit to $\pi\pi$ data.
However, in addition to ChPT,  the model allows for a ``direct''
 contribution of the $I=0$ resonances ($\sigma$ and $f_0$) from the 
vertices $\gamma\gamma \sigma$ and  $\gamma\gamma f_0$ and of the
 $f_2(1270)$ resonance
$\gamma\gamma f_2$:
\begin{itemize}
\item Comparing the predictions on the differential cross-section
 $d\sigma/d\Omega$ with the data on $\gamma\gamma\to\pi^+\pi^-$,
Fig. 11 of \cite{mennessier} shows that the unitarised Born term alone,
 i.e. without any
additional direct contribution, describes, within 20\%, the data in
 \cite{MARK2}. For instance, 
at the peak near threshold, the prediction after angular integration 
is about 480 nb (data 420 nb) while at 800 MeV mass one predicts 150
nb (data 130 nb).
\item Also the data on $d\sigma/d\Omega$ for the process
 $\gamma\gamma\to
\pi^0\pi^0$ \cite{cball} are in rough agreement with the predictions of
 Fig. 14 of \cite{mennessier}. The data (after 20\% acceptance
 correction) indicate $\sim$ 10 nb cross section in the
range 400-800 MeV, which yields $d\sigma/d\Omega\sim 1.7$ nb/str.
The predicted cross section reaches this value at
around 500 MeV and rises to twice this value at 800 MeV. The analysis
indicates, that one does not need a strong direct $\gamma\gamma$
 coupling for describing the data.
\item The situation looks different in the $f_2(1270)$ region. Here the
differential cross section $d\sigma/d\Omega$  at 90$^\circ$ 
is observed in the interval 1200-1300 MeV at $
\sim 50$ nb/str while the prediction from the unitarized 
Born term is around 18
nb/str in Fig. 14 of \cite{mennessier}. Therefore,
the ``direct'' contribution in $f_2\to\gamma\gamma$ should be dominant.
This is consistent with the observation that predictions for the 
radiative decays of tensor mesons involving the ``direct'' 
coupling of photons to the quark
charges are rather successful.

\end{itemize}
\subsection{Estimate of the $\gamma\gamma$ width}
From the model of \cite{mennessier}, we conclude that the major part of
 the
$\gamma\gamma$ cross section below 1 GeV, besides the pion exchange 
Born cross section, is due to pion exchange with rescattering.
The measured cross section of $\gamma\gamma\to\pi^0\pi^0$ 
in the low energy region is then 
related to the same cross section in $\pi\pi$ scattering and assumed to
 be
dominated by the $\sigma$ resonance. 

To estimate the  $\gamma\gamma$
width, we assume that the 
cross section for $\gamma\gamma\to
\pi\pi$ is given by a Breit Wigner form and reads 
(see for example  \cite{POPPE}):
 \bea
 \sigma_{\gamma\gamma\to
\pi\pi}=&&{(2J+1)\over 4}
\ga {8\pi\over q^2}\dr \times\nnb\\
&&\frac{B_{in} B_{out} (M \Gamma_{tot})^2}
  {(M^2-s)^2+(M \Gamma_{tot})^2} ~.
\label{peak}
\eea
Here, the first ratio takes into account the spin states. $B$
is the branching ratio for incoming and outgoing states
$B_{in}=\Gamma_{\gamma\gamma}/\Gamma_{tot}$, $B_{out}=1/3$ for
 $\pi^0\pi^0$,
such that $\Gamma_{\gamma\gamma}$ can be obtained from the cross section at
 the peak position. In general, some background may be present. 


To begin with, we extract for a confirmation of this procedure 
the  $\gamma\gamma$ width of $f_2(1270)$.
Using the Crystal Ball \cite{cball} $\pi^0\pi^0$ and MARK-II $\pi^+\pi^-$ 
data \cite{MARK2} we obtain from (\ref{peak}) $\Gamma_{f_2\to\gamma\gamma}\sim 3.6 $ and
$\sim 2.6$ keV respectively, a range
which compares with 2.6 keV quoted by PDG \cite{PDG}.

\begin{itemize}
\item
We use the same procedure to estimate the $\gamma\gamma$ width 
for the $\sigma$. To relate to the subsequent theoretical predictions, 
as explained in the introduction, we relate 
the mass in Eq. (\ref{peak}) to the ``Breit Wigner mass'' in
 Eq. (\ref{sigmamass}).
In the isoscalar $S$ wave $\pi\pi$ scattering, 
where the phase shift goes through 90$^\circ$, this mass
is around 1 GeV if the effect from $f_0(980)$ is subtracted, and
 decreases to $\sim$850 MeV when including the  $f_0(980)$ and $f_2$. 
In the following, we  shall use here the range: 
\beq
M_\sigma\simeq (750\sim 1000)~{\rm  MeV}~.
\label{bwmass}
\eeq
While we assume 
there is no background under the resonance in the $I=0$ channel we
include the slowly varying $I=2$ 
amplitude as in $\pi^+\pi^-\to \pi^0\pi^0$ given in Eq. (\ref{amp00}).
At 900 MeV the $I=2$ phase reaches $\delta_2\approx
-20^\circ$.  This ``background'' reduces the peak position by the factor
0.88, which is a small correction as, a priori, expected.
We obtained a good description of the Crystal Ball
$\gamma\gamma\to \pi^0\pi^0$ cross section data \cite{cball}
in the energy range 350-800 MeV before D wave scattering becomes
 important
by using the form (\ref{peak}) with $J=0$ and
a mass dependent width, including the phase space factor:
\beq
\Gamma_{tot}(s)=\Gamma_\sigma \sqrt{ { {s-4m^2_\pi}\over
 {M^2_\sigma-4m^2_\pi}}}~,
\eeq
for the parameters
$M_\sigma$ considered.
The branching ratio $B_{\gamma\gamma}$ is then obtained from the peak
 cross
section $\sigma_{peak}$ at $\sqrt{s}\to M_\sigma$ independently from
 $\Gamma_{tot}$ and
its mass dependence. 

We see that, around 800 MeV,
the cross section for the process is around 10 nb after acceptance
correction. At higher masses, the S wave cross section is hidden under
 an
increasing $f_2(1270)$ contribution, where 
we take again 10 nb by extrapolation. 
Then for the considered range of
Breit Wigner masses $M_\sigma$ in Eq. (\ref{bwmass}) and
 $\Gamma_{\sigma,tot}\simeq M_\sigma$, we find:
\begin{equation}
\Gamma(\sigma\to\gamma\gamma)
   \approx (1.4 \sim 3.2)~{\rm  keV}~,   
\label{sigmawidth}
\end{equation} 
which is  to be considered as an estimate of the full radiative width
 \footnote{In \cite{mennessier}, a larger value of the full radiative 
width of 5
 to 9 keV has been 
obtained in order to fit the DM2 and some older data \cite{DM2} which
 were two times bigger than the one used in this paper.}. 
\item 
Aiming at a further theoretical interpretation, we decompose the amplitude
for $\gamma\gamma\to \sigma\to \pi \pi$ into two components
$T_\sigma = T_{direct}+T_{resc}$ 
 using the model of \cite{mennessier}. Theoretical models on glueball decays
usually do not include $T_{resc}$ but refer to $T_{dir}$.
An upper limit of the direct coupling:  
\begin{equation}
\Gamma(\sigma\to\gamma\gamma)|_{direct} \leq 1.4 ~{\rm keV}~,
\label{sigmabound}
\end{equation}
can be obtained in the case of a negative interference of the 
re-scattering amplitude shown in \cite{mennessier} 
for $\gamma \gamma \to \pi^0 \pi^0$ with the direct amplitude.
With our fit of the model to the $\pi^+\pi^-$ and $\pi^0 \pi^0$ 
data involving 
both the $\pi\pi$ rescattering and direct meson coupling we obtain a
 small $\gamma\gamma$ width compatible with the previous bound:
\begin{equation}
\Gamma(\sigma\to\gamma\gamma)|_{direct} \approx {\rm 0.3~{\rm keV}}~,
\label{sigmadirect}
\end{equation}
within an accuracy of about 50\%. This result corresponds to the value
 of the coupling $|f|\approx 0.1$ defined in \cite{mennessier}. In
 an earlier analysis, a value $f\simeq 0.65$ corresponding to a
 $\gamma\gamma$ width of 6 keV and for $M_\sigma\simeq  600$ MeV has been found
  \cite{mennessier}, which   would correspond to data two times bigger
 \cite{DM2}. The accuracy of this estimate can be improved using more accurate data below 600 MeV.
\item One can notice that, the highest value of the width in Eq.
(\ref{sigmawidth}), corresponding to the highest mass $M_\sigma$, 
is comparable to the width in Eq. (\ref{widthpenn}) 
\cite{PENNINGTON}.  However, no separation into different contributions has
been considered in that analysis.
\end{itemize}
What this result indicates is, that using 
the wide and heavy $\sigma$ along the
physical region, yields a similar result than working with 
the complex pole at low energies in (\ref{sigma}).
Only, if one used a low energy mass around 440 MeV and took the
corresponding Breit Wigner cross section, 
would one get the much smaller radiative
 width
an order of magnitude smaller, but this would apparently be an
 inconsistent
procedure.
\section{Comparison with QSSR predictions}\label{qssr}
\begin{itemize}
\item The QSSR determinations of the $\bar q q$ and gluonium light
 scalar meson
masses are performed in a narrow width approximation \footnote{A
 Breit-Wigner parametrization of the spectral function leads to a tiny
 width correction on the mass prediction \cite{VENEZIA,SNG,SNB}.}, 
i.e. with a real pole.
The predicted value in Eq. (\ref{eq:scalarmass})
is compatible with the observed properties
of the ``visible'' meson having a Breit-Wigner mass with parameters
 given in Eq. (\ref{bwmass}),
which is closer to the theoretical calculation than
the complex mass of the $\sigma$ meson in Eq. (\ref{sigma}) with large
 imaginary
part. 
\item A radiative decay width of the size in Eq. (\ref{sigmadirect}) is
 expected from a direct bare (index B) unmixed gluonium $\sigma_B$ decays
 obtained from QSSR \cite{VENEZIA,SNG} :
\beq
\Gamma (\sigma_B\rar\gamma\gamma)\simeq (0.2\sim 0.3)~{\rm keV}.
\eeq
A width of this size induces a tiny effect of about $3 \times 10^{-11}$ 
to the
 muon anomalous magnetic moment \cite{SNANOM}. \\
One can also notice that,
 within the QSSR approach, a four-quark state gives a much smaller width
 of the order of $10^{-3}$ keV \cite{SNA0}, while a $\bar qq$ state
 leads to a larger width~\cite{SNA0,BN}.
\item A determination of the total hadronic width
using, either a dispersion representation of the scalar-pion-pion
 vertex, or a Breit-Wigner form of the $\sigma_B$ in the two-point
 function sum rule, leads to a value  \cite{VENEZIA,SNG,SN06}:
\beq
\Gamma_{\sigma_B \to\pi\pi}\approx  {3\over 2}(200\sim 700) ~{\rm MeV}~,
\eeq
in agreement with the ones from $\pi\pi\to \pi\pi$ and
 $\pi\pi\to KK$ scattering data \cite{mennessier}. It also follows that a
 $\sigma$ having a Breit-Wigner mass below 750 MeV cannot be wide  \cite{VENEZIA,SNG,SN06} (see also some papers in Ref. \cite{SNG1}) . Up to $SU(3)$ breaking corrections, one also gets the relation:
\beq
g_{\sigma_B\pi\pi}\approx \sqrt{3\over 4}g_{\sigma_B KK}~.
\eeq
One should remark that the strong coupling of the $\sigma$ to $\pi\pi$ and $\bar KK$ is a characteristic
gluonium ($\bar qq$ singlet) feature which is not present for a four-quark state or $\bar KK$ molecule model for the $\sigma$.  A careful measurement of such couplings may select among the different scalar meson scenarios. 
\end{itemize}
\section{Summary and conclusions}
\begin{itemize}
\item We have reanalyzed  the $\gamma\gamma$ scattering data and
 concluded that in the
mass region below 1 GeV the cross section for 
$\gamma\gamma\to \pi^0\pi^0$ can be largely explained by
the one pion exchange process with $ \pi\pi$ rescattering.  An improvement
of our estimates needs more accurate data below 600 MeV. 
\item The small direct coupling of the $\sigma$ to $\gamma\gamma$ and
 its large hadronic width are consistent
with a large gluonic component of  the $ \sigma$ resonance, expected
 from QSSR calculations (see section \ref{qssr}).  
\item  The large gluonic component 
of the $\sigma$ has been exploited in some
 phenomenological models with glueball and $\bar qq$ nonet \cite{OCHS}
and with a maximal $\bar qq$-gluonium mixing below 1 GeV \cite{BN,SNG,SN06}.
 We plan to come back to these different mixing models  and to analyze
 the
nature of some other scalar mesons in a future work.
\item In addition to the present analysis of $\gamma\gamma$ and
 $\pi\pi$ scattering data, some tests of the gluon component of the $\sigma$
 have been proposed in the literature, like e.g. the one from $D$
 semi-leptonic decays \cite{DOSCH} where in addition to $\pi\pi$ one equally
 also expects the one into $\bar KK$. $B\to K\bar KK $ is also expected
 to be a source of gluonic decay from the $b\to sg$ process \cite{minkB}.
  
\end{itemize}

\section*{Acknowledgements}
One of us (W.O.) would like to thank Stephan Narison for the invitation and for the kind hospitality at the
Montpellier University within the grant provided by
Region of Languedoc-Roussillon Septimanie. 


\begin{thebibliography}{999}
\bibitem{MONTANET} For reviews, see e.g.:
L. Montanet, {\it Nucl.
 Phys. Proc. Suppl.} {\bf 86} (2000) 381; talk given at the Gribov Memory,
 WSC 2001; U. Gastaldi, Rencontres de Physique de la Vall\'ee d'Aoste,
 27 Feb-4 March  (2000). 
\bibitem{MORGAN} D. Morgan,  {\it Phys. Lett.} {\bf B 51} (1974) 71.

\bibitem{BN} A. Bramon and S. Narison, {\it Mod. Phys. Lett.} {\bf A 4}
 (1989) 1113.
\bibitem{SNG} S. Narison, {\it Nucl. Phys.} {\bf B 509} (1998) 312;  S.
 Narison, {\it Nucl. Phys. Proc. Suppl.} {\bf 64} (1998) 210; S. Narison, {\it Nucl. Phys.
 Proc. Suppl.} {\bf 96} (2001) 244.
\bibitem{KLEMPT} 
E. Klempt, B.C. Metsch, C.R. M$\ddot {\rm u}$nz and H.R. Petry,
{\it Phys. Lett.} {\bf B 361} (1995) 160.
\bibitem{OCHS}

P. Minkowski and W. Ochs, {\it  Eur. Phys. J.} {\bf  C
 9} (1999) 283.

\bibitem{SN06} S. Narison, {\it Phys. Rev.} {\bf D 73} (2006) 114024.
\bibitem{TORN}N. Tornqvist, {\it Phys. Rev. Lett.} {\bf 49} (1982) 624;
 {\it Z. Phys.} {\bf C 68} (1995) 467.
\bibitem{BEVEREN} E. van Beveren et al., {\it Z. Phys.} {\bf C 30} (1986)
615;  E. van Beveren and G. Rupp,  {\it Eur.Phys.J.} {\bf C 10} (1999) 469.
\bibitem{JAFFE1}R. Jaffe, {\it Phys. Rev.} {\bf D 15} (1977) 267;  {\it
 Phys. Rev.} {\bf D 15} (1977) 281.
\bibitem{BLACK}J.M. Richard,  {\it Nucl. Phys. Proc. Suppl.} {\bf 164} (2005) 131;
D. Black et al., {\it Phys. Rev.} {\bf D 59} (1999)
 074026; A. Fariborz, J. Schechter, {\it Phys. Rev.} {\bf D 60} (1999)
 034002; D. Wong and K.F. Liu, {\it Phys. Rev.} {\bf D 21} (1980) 2039; 
J.I. Latorre and P. Pascual, {\it Jour. Phys. } {\bf G 11} (1985) L231; 
M. Alford and R.L. Jaffe, {\it Nucl. Phys.} {\bf B 509} (1998) 312; 
F. Buccella et al., {\it Eur. Phys. J.} {\bf C~49} (2007) 743;
M. Karliner, H. J. Lipkin, {\it Phys. Lett.} {\bf B 612} (2005) 197;
L. Maiani et al., {\it Phys. Rev. Lett.} {\bf  93} (2004) 212002;
A. Selem and F.Wilczek, Proc. Ringberg workshop, "New Trends in HERA 
Physics 2005"[hep-ph/0602128];
T.V. Brito et al., {\it Phys. Lett.} {\bf  B 608} (2005) 69; R.D. Matheus et al., hep-ph/0705.1357.
\bibitem{SNA0} S. Narison, {\it Phys. Lett.} {\bf B 175} (1986) 88.
\bibitem{ACHASOV}N.N. Achasov, S.A. Devyanin and G.N. Shestakov, {\it
 Z. Phys.} {\bf C 16} (1984) 55; C. Hanhart et al., hep-ph / 0701214.
\bibitem{ISGUR}N. Isgur and J. Weinstein, {\it Phys. Rev.} {\bf D 41}
 (1990) 2236.
\bibitem{BARNES} T. Barnes,  {\it Phys. Lett.} {\bf B 165} (1985) 434;
 Proc. IXth Int. Workshop on $\gamma\gamma$ collisions, World Scientific
 (1992) 263 ed. D. Caldwell and H.P. Paar.
\bibitem{HOLINDE}G. Janssen et al., {\it Phys. Rev.} {\bf D 52} (1995)
 2690 and references therein.
\bibitem{NSVZ}V.A. Novikov et al., {\it Nucl. Phys.} {\bf B 191} (1981)
 301; K. Chetyrkin, S. Narison and V.I. Zakharov, {\it Nucl. Phys.} {\bf B 550} (1999)
353.
\bibitem{SNG0}S. Narison, {\it Z. Phys.} {\bf  C 22} (1984) 161.
\bibitem{SNG1} P. Pascual and R. Tarrach, {\it Phys. Lett.} {\bf B 113} (1982) 495;
S. Narison, {\it Phys. Lett.} {\bf B 125} (1983) 501;
C.A. Dominguez and N. Paver,   {\it Z. Phys.} {\bf  C 31} (1986) 591;
J. Bordes, V. Gimenez and J.A. Pe\~narrocha,  {\it Phys. Lett.} {\bf B 223} (1989) 251;
E. Bagan and T.G.  Steele,  {\it Phys. Lett.} {\bf B 243} (1990) 413; 
J.L. Liu and D. Liu, {\it J. Phys} {\bf G 19} (1993) 373; L.S. Kisslinger, J. Gardner
and C. Vanderstraeten,  {\it Phys. Lett.} {\bf B 410} (1997) 1; T. Huang, H.Y. Jin and A.L.  Zhang,
 {\it Phys. Rev.} {\bf D 58} (1998) 312;
T.G. Steele, D. Harnett and G. Orlandini, {\it AIP Conf. Proc.} {\bf  688} (2004) [hep-ph/0308074]; H. Forkel, {\it Phys. Rev.} {\bf D 71} (2005) 054008. 
\bibitem{VENEZIA} S. Narison and G. Veneziano, {\it Int. J. Mod. Phys.}
 {\bf A 4, 11} (1989) 2751.
\bibitem{PEARDON}  C.  Morningstar and  M.  J.   Peardon ,
 {\it Phys. Rev.} {\bf D 60} (1999) 034509;
  G. Bali et al., UKQCD Collaboration ,  {\it Phys. Rev.}  {\bf D 62} (2000)
 054503;  A. Hart and M. Teper, {\it Phys. Rev.}  {\bf D 65} (2002) 34502; Y.
 Chen et al., {\it Phys. Rev.} {\bf D 74} (2006) 094005; ; H. Wada et
 al. , hep-lat/0702023.
 \bibitem{MICHAEL}A. Hart, C. McNeile, C. Michael and J. Pickavance, 
  {\it Phys. Rev.} {\bf D 74} (2006) 114504; 
T. Kunihiro et al., SCALAR Collaboration, {\it Phys. Rev.} {\bf D
 70} (2004) 034504. 
\bibitem{CLOSE} C. Amsler and F.E. Close, {\it Phys. Rev.} {\bf D 53}
 (1996) 295;
D. Weingarten, {\it Nucl. Phys. Proc. Suppl.} {\bf 73}
 (1999) 283; 
F.E. Close and A. Kirk, {\it Eur.Phys.J.} {\bf C 21} (2001) 531;
F. Giacosa, Th. Gutsche, V.E. Lyubovitskij and A. Faessler,
 {\it Phys. Lett.} {\bf B 622} (2005) 277;
X.G. He, X.Q. Li, X. Liu and X.Q. Zeng, {\it Phys. Rev.} {\bf D
 73} (2006) 114026.
\bibitem{aniso}V. Anisovich, Yu. Prokoshkin and A. Sarantsev, 
 {\it Phys. Lett.} {\bf B 389} (1996) 388.
\bibitem{LANIK} P. Di Vecchia and G. Veneziano, {\it Nucl. Phys.} {\bf
 B 171} (1980) 253;  J. Ellis and J. Lanik,  {\it Phys. Lett.} {\bf B
 150} (1985) 289;  {\it Phys. Lett.} {\bf B 175} (1986) 83; P. Jain, R.
 Johnson and J. Schechter, 
{\it Phys. Rev.} {\bf D 35} (1987) 2230.
\bibitem{ZHENG}L. Y. Xiao, H. Q. Zheng and Z. Y. Zhou, QCD 06
 (Montpellier 2006) and references therein.
\bibitem{PICH} G. Ecker et al., {\it Nucl. Phys.} {\bf B321} (1989) 31.
\bibitem{MIN} H. Fritzsch and P. Minkowski, {\it Nuovo Cimento} {\bf A
 30} (1975) 393.
\bibitem{SVZ}M.A. Shifman, A.I. Vainshtein and V.I. Zakharov,
{\it
 Nucl. Phys.} {\bf B 147} (1979) 385, 448.
\bibitem{SNB}For reviews, see e.g.: S. Narison, {\it QCD as a theory of
 hadrons,
Cambridge Monogr. Part. Phys. Nucl. Phys. Cosmol.} {\bf 17}
 (2004) 1-778 [hep-ph/0205006]; S. Narison, {\it QCD
spectral sum rules ,
  World Sci. Lect. Notes Phys.} {\bf 26} (1989) 1-527; S. Narison,
{\it
 Acta Phys. Pol.} {\bf 26} (1995) 687; S. Narison, {\it Riv. Nuov.
 Cim.} {\bf 10 N2} (1987) 1; S. Narison, {\it Phys. Rep.}  {\bf 84} (1982)
 and references therein. 
\bibitem{CERN-MUNICH} B. Hyams et al., {\it Nucl. Phys.} {\bf B 64}
 (1973) 134;
G. Grayer et al.,  {\it Nucl. Phys.} {\bf B 75} (1974) 189.

\bibitem{EM}
P. Estabrooks and A.D. Martin, Nucl. Phys. B95 (1975) 322.
\bibitem{CM2}
B. Hyams et al., Nucl. Phys. B100 (1975) 205. 

\bibitem{KPY}
R. Kami\'nski, J.R. Pel\'aez and F.J. Yndur\'ain (2006),
 {\it Phys. Rev.} {\bf D 74} (2006) 014001, Erratum ibid. {\bf D 74} (2006)
079903.

\bibitem{ochs06}  W. Ochs, Contribution to QCD06,
July3-7,2006, Montpellier,   
France, to be publ. in {\it Nucl. Phys. Proc. Suppl.}
 [hep-ph/0609207].

\bibitem{GAMS}  
D. Alde et al., Eur. Phys. J. A3 (1998) 361.

\bibitem{ESTABROOKS} P. Estabrooks,  {\it Phys. Rev.} {\bf D 19} (1979)
2678. 
\bibitem{AMP} K.L. Au, D. Morgan and M. Pennington, 
    {\it Phys. Rev.} {\bf D 35} (1987) 1633.
\bibitem{mennessier}
G. Mennessier, {\it Z. Phys.} {\bf  C 16} (1983) 241; G. Mennessier,
 Montpellier preprint PM/81/6 (unpublished);
O. Babelon et al., {\it Nucl. Phys.} {\bf B 113} (1976) 445; 
G. Mennessier and T.N. Truong, {\it Phys. Lett.} {\bf B 177} (1986) 195.
A Pean, Th\` ese doctorat Montpellier 1992 (unpublished).

\bibitem{mo2}
P. Minkowski and W. Ochs, 
{\it Nucl. Phys. Proc. Suppl.} {\bf 121} (2003) 119; {\it
 Nucl. Phys. Proc. Suppl.} {\bf 121} (2003) 121.

\bibitem{minkB} P. Minkowski and W. Ochs, {\it  Eur. Phys. J.} {\bf  C
 39} (2005) 71.

\bibitem{PENNINGTON}M.R. Pennington, hep-ph/0604212 and
 {\it Mod. Phys. Lett.} {\bf A 22} (2007) 1439; 
M. Boglione and M.R. Pennington, {\it Eur.Phys.J.} {\bf C9} (1999) 11.
\bibitem{leutwyler} I. Caprini, G. Colangelo and H. Leutwyler,  {\it
  Phys. Rev. Lett.} {\bf 96} (2006) 132001.
\bibitem{YND} F.J. Yndurain , R. Garcia-Martin and J.R. Pelaez,
 hep-ph/0701025 and references therein.
\bibitem{donoghue} J.F. Donoghue, B.R. Holstein and Y.C. Lin, {\it
  Phys. Rev.} {\bf D 37} (1988) 2423; 
J.F. Donoghue and  B.R. Holstein, {\it  Phys. Rev.} {\bf D 48} (1993)
 137; J. Bijnens and F. Cornet, {\it Nucl. Phys.} {\bf B 296} (1988) 557;
 L.V. Fil'kov, V.L. Kashevarov,  {\it  Phys. Rev.} {\bf C 73} (2006)
 035210.
\bibitem{cball} H. Marsiske et al.,  Crystal Ball collaboration,
 {\it Phys. Rev.} {\bf  D 41} (1990) 3324.
\bibitem{bellucci} S. Bellucci, J. Gasser and M.E. Sainio,
 {\it Nucl. Phys.} {\bf B 423} (1994) 80, Erratum-ibid. {\bf B431} (1994) 413;  
M. Knecht, B. Moussallam, J. Stern {\it Nucl. Phys.} {\bf B 429} (1994) 125.
\bibitem{MARK2}  J. Boyer et al., MARK II Collaboration, {\it Phys.
 Rev.}
{\bf D42} (1990) 1350.
\bibitem{POPPE}G. Poppe, {\it Int. J. Mod. Phys.}  {\bf A1} (1986) 545.
\bibitem{PDG}The Particle Data Group, W.M. Yao et al,  {\it Jour.
 Phys.} {\bf G 33} (2006) 1.
\bibitem{DM2}A. Coureau et al., DM2 collaboration, {\it Phys.
 Lett.} {\bf B 96} (1980) 402; 
Ch. Berger et al., PLUTO collaboration, {\it Phys. Lett.} {\bf B
 94} (1980) 254; 
R. Brandelik et al., TASSO collaboration, {\it Z. Phys.} {\bf  C
 10} (1981) 117.
\bibitem{SNANOM}S. Narison, {\it Phys. Lett.} {\bf B 568} (2003) 231.
\bibitem{DOSCH}H. G. Dosch and S. Narison, {\it Nucl. Phys. Proc.
 Suppl.} {\bf 121} (2003) 114.
\end{thebibliography}
\end{document}